\documentclass[preprint2]{aastex}

\slugcomment{Astronomical Journal - Pre-referee version.}

\begin{document}

\title{Study of the interacting system NGC\,6845}

\author{Irapuan Rodrigues\altaffilmark{1,2}, Horacio
  Dottori\altaffilmark{2},} \affil{Dep. de Astronomia, Instituto de
  F\'{\i}sica, UFRGS, CxP 15051, CEP 91501-907 Porto Alegre, Brazil}
\email{irapuan@if.ufrgs.br}

\author{Elias Brinks} \affil{Univ. de Guanajuato, Apartado Postal 144,
  Guanajuato, C.P.  36000, Gto, M\'exico}
 
\and

\author{I. Felix Mirabel} \affil{CEA, DSM, DAPNIA, Service
  d'Astrophysique, Centre d'Etudes de Saclay, 91191 Gif--sur--Yvette
  Cedex, France; and IAFE, cc 67, snc 28 (1948), Buenos Aires, Argentina}

\altaffiltext{1}{Brazilian CNPq fellow.} 

\altaffiltext{2}{Visiting astronomer, Complejo Astron\'omico El
  Leoncito, operated under agreement between the Consejo Nacional de
  Investigaciones Cient\'{\i}ficas y T\'ecnicas de la Rep\'ublica
  Argentina and the Universities of La Plata, C\'ordoba y San Juan.}

%%%%%%%%%%%%%%%%
\begin{abstract}
  We present optical spectroscopy, B,V,R and I CCD photometry and VLA
  neutral hydrogen observations of the interacting quartet NGC\,6845,
  also know as Klemola\,30. NGC\,6845\,A, the dominant component,
  sports a broad and bright tidal bridge and a faint tidal tail, which
  bifurcate. The tidal bridge has a (B-I) color bluer than that of
  NGC\,6845\,A inner disk. Five strong condensations, identified as
  \mbox{\rm H{\sc ii}} regions brighter than the brightest one known
  in our Galaxy, are found along the tidal bridge, with the two most
  luminous located at the bridge tip. Two giant \mbox{\rm H{\sc ii}}
  regions, comparable to 30\,Dor, are located where the tidal bridge
  and the tidal tail joint the disk of NGC\,6845\,A. Since the age of
  the \mbox{\rm H{\sc ii}} regions is 3-8\,Myr, star formation has
  been occurring along the tidal bridge and in the tidal arm well
  after they had begun to be torn apart ($\geq$100\,Myr).  Satoh model
  fitting to the optical rotation curve of the A component reveals a
  kinematical mass of 4.4\,$(\pm1.2)\,10^{11}\,M\odot$, inside the
  central 12\,kpc (H$_0=75{\rm{\,km\,s^{-1}\,Mpc^{-1}}}$).  The
  \mbox{\rm H{\sc i}} emission consist of two components, a more
  massive one that belongs to NGC\,6845\,A, and a second one
  associated to NGC\,6845\,B. We do not detect gas associated to
  galaxies C and D.  The total amount of \mbox{\rm H{\sc i}} is
  1.4$\times\,10^{10}M\odot$, five time higher than the \mbox{\rm
    H{\sc i}} content of the Milky Way. The HI kinematics indicates an
  amount of dark matter associated to the A component two times higher
  than the mass inside its central 12\,kpc. The group kinematics
  indicates an ${\rm M/L}\approx 43(\pm 2)\,{\rm (M/L)}_\odot$ or
  ${\rm M/L}\approx 66(\pm 2)\,{\rm (M/L)}_\odot$, according to two
  different prescriptions for the internal absorption correction. In
  spite of this large difference, both values furnish similar group
  mass ($\approx 1 \times 10^{13}\,{\rm M_\odot}$).  Although
  preliminary results on N-Body simulations indicate that either B or
  C might well create a tidal feature like the bridge of the A
  component, the collision with B appears to be more favourable.
\end{abstract}


%%%%%%%%%%%%%%%%
\section{Introduction \label{intro}}
%%%%%%%%%%%%%%%%

Two decades ago, \citet{tt} demonstrated through numerical models that
gravitational interactions between galaxies can generate spectacular
tidal features. In earlier work, \citet{zwicky} already proposed that
collisions between giant galaxies may result in the ejection of stars
and gas into the intergalactic space, resulting in the formation of
star clusters and star forming dwarf galaxies.  Direct observational
evidence for the existence of such a class of recycled galaxies has
already been reported in the literature~\citep[e.g.][]{schweizer}.
Systems like the Antennae~\citep{schweizer,mdl} and the
Super--antennae~\citep{mlm} show, in addition to stellar material and
gas torn from the disks of the interacting galaxies, optically bright
clumps along and at the extremes of the tidal tails as well. In some
cases there is even a chain of knots of bright material whose colors
become bluer towards the tips~\citep{hggs}. Radio observations of
these systems have shown that a large amount of atomic hydrogen is
torn away from the parent galaxies, stretched along the tidal tails,
and seems to concentrate at the tips.  In contrast, molecular hydrogen
is concentrated in the central regions of the parent galaxies. A
particularly nice example is the interacting galaxy
Arp\,105~\citep{dm,dbwm}, that shows near the end of its tidal tail a
LMC--like galaxy, resulting from the collision of a spiral with an
elliptical galaxy.

It is well known from numerical simulations that the form and strength
of the tidal bridges and tails depend on the orbital and spin
parameters of the colliding galaxies
\citep[e.g.][]{tt,wright,eneev,nw,barnes88}.  However, the star
formation activity within the tidal tails could only recently be
numerically studied, after the development of appropriate codes
\citep[e.g.][]{hk,ekt}. Models that do not take into account the gas,
or the fact that the \mbox{\rm H{\sc i}} is distributed predominantly
at the periphery of spirals, are unable to reproduce the observed
distribution after a collision \citep{bh}, while those that do take
this into account find that, after a collision with a massive body,
the gas can form huge pools at the extreme ends of the tidal tails
\citep{ekt}.

In this paper we study the group NGC\,6845 (Klemola\,30,
AM\,1957-471). \citet{klem} described the group as being compact,
dominated by a bright, one--armed spiral. In the \citet{am} catalogue,
NGC\,6845 is described as an interacting quartet.  Its four components
are approximately aligned in a SW--NE direction.  The components
NGC\,6845\,A and NGC\,6845\,B, the later located to the NE, show
abundant signs of an ongoing interaction, such as tidal bridges and
tails. Many clearly defined knots dot the tidal bridge that links
NGC\,6845\,A to NGC\,6845\,B. Signs of interaction seem also to be
present in NGC\,6845\,C and NGC\,6845\,D, according to \citet{gr} and
\citet{rosegraham} sketches.  In order to study the collision
kinematics and the subsequent process of star formation, we carried
out optical imaging and spectroscopy as well as radio synthesis
observations.

In Section~\ref{sec:obs} we show our optical and radio observations of
NGC\,6845 and briefly describe the data reduction; in
Sec.~\ref{sec:res} we present our analysis; two possible scenarios for
the interaction process are discussed in Sec.~\ref{sec:dynamics} and
in Sec.~\ref{sec:conclu} we summarize our results and present our
conclusions.

%%%%%%%%%%%%%%%%
\section{Observations and data reduction\label{sec:obs}}
%%%%%%%%%%%%%%%%

Table~\ref{tab:gal} presents some basic data on NGC\,6845. Our
observations consist of optical broadband photometry and spectroscopy
obtained at the CASLEO\footnote{Complejo Astron\'omico El Leoncito is
  operated under agreement between the Consejo Nacional de
  Investigaciones Cient\'{\i}ficas y T\'ecnicas de la Rep\'ublica
  Argentina and the Universities of La Plata, C\'ordoba y San Juan.}
and Bosque Alegre\footnote{Estaci\'on Astrof\'{\i}sica de Bosque
  Alegre is operated by Observatorio Astron\'omico of the Universidad
  Nacional de C\'ordoba, Argentina.} observatories and \mbox{\rm H{\sc
    i}} observations taken with the Very Large Array.\footnote{The VLA
  is a facility of the National Radio Astronomy Observatory which is
  operated by Associated Universities, Inc., under cooperative
  agreement with the National Science Foundation.}
Tables~\ref{tab:ima} and~\ref{tab:spec} list the relevant instrumental
and observational parameters. We adopted
H$_0=75{\rm{\,km\,s^{-1}\,Mpc^{-1}}}$ thoughout this paper, which
results in a distance to NGC\,6845 of 85\,Mpc. At this distance, 1
arcsec corresponds to 412\,pc. The observations are described in more
detail below.

%%%%%%%%%%%%%%%%
\subsection{Optical images\label{subsec:otico}}
%%%%%%%%%%%%%%%%
Broad--band B, V, R and I CCD images were obtained in September 1997
with the Ritchey--Chr\'etien 2.15--m telescope at CASLEO, Argentina,
using a thinned Tektronix $1024\times1024$ pixel CCD detector. The
telescope focal ratio is f/8. The final image scale, with a focal
reducer, is $0\farcs 81\,{\rm pixel}^{-1}$. The seeing during the
observations varied between 2\arcsec\ and 4\farcs 5.  Data reduction
was performed with IRAF, using standard procedures. Different frames
taken through the same filter were registered to a common center,
using field stars as a reference, and co--added to remove cosmic ray
events and increase the dynamic range. Small seeing differences among
frames were corrected by gaussian filtering. Flux calibration was
based on standard stars from \citet{lan}. We corrected as usual for
atmospheric extinction.

%%%%%%%%%%%%%%%%
\subsection{Optical spectroscopy}\label{subsec:spec}
%%%%%%%%%%%%%%%%

Long--slit spectroscopic observations were carried out on July 1994
(seeing\,=\,1\farcs 7) and September 1997 (seeing\,=\,2\arcsec) with
the 2.15\,m telescope of the CASLEO observatory, and on December 1997
with the Bosque Alegre 1.52\,m telescope (seeing\,=\,1\farcs 5). In
Table~\ref{tab:spec} we present the log of the spectroscopic
observations. The slit positions, given in the first column of this
table, are marked in Figure~\ref{fig:slit}.

Data processing was again done with IRAF following standard methods
for bias subtraction, flat--fielding, and image combination. The
resulting two dimensional spectra were wavelength calibrated using
\mbox{HeNeAr} comparison spectra and flux calibrated using
spectrophotometric standard stars from \citet{sb}.  Galactic
extinction was corrected with the standard absorption law, adopting
$A_B=0.16$ \citep{buh}.

%%%%%%%%%%%%%%%%%%%%%%%%%%%%%%%%%
\subsection{VLA radio--observations\label{subsec:vla}}
%%%%%%%%%%%%%%%%%%%%%%%%%%%%%%%%%
VLA D--array observations of the NGC\,6845 system were obtained during
a short, 1.5 hour run in 1996, June 19. Due to its far southern
declination the synthesized beam is substantially elongated in the
N--S direction. The beam measures $112\arcsec\times37\arcsec$, at a
$+14.5\degr$ angle measured from North to East. The sensitivity of the
system was rather modest and the \mbox{\rm H{\sc i}} data should be
seen as exploratory. As it was not obvious at what velocities
\mbox{\rm H{\sc i}} could be expected we aimed for velocity coverage
rather than for velocity resolution, trying to maximize our
sensitivity for the velocity range expected for objects A and B
(radial velocities are given in Tab.~\ref{tab:gal}). The velocity
resolution was $43.1{\rm{\,km\,s}^{-1}}$, covering the interval
$6450{\rm{\,km\,s}^{-1}}$ to $6900{\rm{\,km\,s}^{-1}}$ with both
right--hand (R) and left--hand (L) polarization; the intervals from
$6000{\rm{\,km\,s}^{-1}}$ to $6450{\rm{\,km\,s}^{-1}}$ and
$6900{\rm{\,km\,s}^{-1}}$ to $7050{\rm{\,km\,s}^{-1}}$, were observed
with one polarization only. Data calibration and mapping was performed
with AIPS. The final maps have an RMS noise of
$1.6{\rm{\,mJy\,beam}^{-1}}$ or 0.23\,Kelvin (where
$1{\rm{\,mJy\,beam}^{-1}}$ corresponds to 0.15\,K).

%%%%%%%%%%%%%%%%
\section{Results\label{sec:res}}
%%%%%%%%%%%%%%%%

%%%%%%%%%%%%%%%%
\subsection{Optical morphology and photometry\label{subsec:phot}}
%%%%%%%%%%%%%%%%

We present in Figure~\ref{fig:rimag} the R image, enhancing the lowest
surface brightness levels in order to highlight the faint details that
might have been produced by the interaction. Clear indications of
strong collision are observed in NGC\,6845\,A and NGC\,6845\,B. The
tidal bridge between galaxies A and B present a fainter counterpart, a
tidal tail that bifurcate and seems to connect to components C and D
(see sketches in \citealt{gr} and \citealt{rosegraham} and Plate I
in \citealt{rose}).

In Table~\ref{tab:phot_gal} we present the integrated B magnitude and
colors (B-V), (V-R) and (V-I) of the group members. As can be seen
from Tab.~\ref{tab:gal}, our B magnitudes are in better agreement with
\citet{rosegraham} than with \citeauthor{rc3}. The (B-I) color map
(Fig.~\ref{fig:colorBI}) shows that the color of the tidal bridge is
bluer than the inner disk of A and B components and comparable to
their outer disks. \citet{hggs} pointed out that in the case of the
evolved merge NGC\,7252 the tidal tails are 0.2--0.5 mag bluer in
(B-R) than the outer regions of the main body.

NGC\,6845\,A is classified as SB(s)b (\citeauthor{rc3}). The color
distribution along its bar is not uniform. To the SW, the bar is much
redder, indicating that on this side we are seeing dust in front of
the stellar bar, where it is usually accumulated in more or less
linear shock fronts. The bar of NGC\,6845\,A is shifted by $30\degr$
counterclockwise with respect to the photometric major axis, and it
may have been tidally created by the encounter, as suggested for
example by \citet{barnes88} or \citeauthor{salo}'s~\citeyearpar{salo}
simulations of disk-disk encounters.

Plumes and small condensations are evident around NGC\,6845\,B, and
are likely debris of the tidal interaction. The tidal bridge linking
the NGC\,6845\,A and B galaxies is very pronounced and knotty
(Fig.~\ref{fig:zoomvel}). The knots are \mbox{\rm H{\sc ii}} regions,
as will be shown in Sec.~\ref{subsec:knots}.

\citeauthor{rose}'s (1979) CTIO 4\,m plate shows more clearly than our
Fig.~\ref{fig:rimag} the presence of condensations in the tidal tail,
specially a set of them close to the NW side of NGC\,6845\,C, though
it does not shows the bridge knotty structure because it is
over--exposed. Based on \citeauthor{rose}'s plate, \citet{rosegraham}
schematically suggested that NGC\,6845\,A western tidal tail
bifurcates and links to NGC\,6845\,C and D galaxies.

%%%%%%%%%%%%%%%%
\subsection{Knots, their brightness and colors\label{subsec:knots}}
%%%%%%%%%%%%%%%%

Table~\ref{tab:phot_reg} shows the V magnitude and (B-V), (V-R) and
(V-I) color indexes for the most conspicuous knots marked in
Fig~\ref{fig:zoomvel}. Without exception the knots present an
\mbox{\rm H{\sc ii}}--region like spectrum, as shown in
Figure~\ref{fig:spectra}, and we feel therefore confident identifying
them as regions of recent star formation. In order to correct the
colors for internal absorption we used for all regions the mean value
of the $\mbox{\rm H$\alpha$} / \mbox{\rm H$\beta$}$ ratio as was
measured in knots 4, 7 and 8 (see Tab.~\ref{tab:flux}). \mbox{\rm
  H$\alpha$} and \mbox{\rm H$\beta$} intensities were measured from
slit positions 3 and 5 for knot 4 and slit position 4 for knots 7 and
8 (see Fig.~\ref{fig:slit}). Although these regions are located at
very different positions within the system, they present roughly the
same $\mbox{\rm H$\alpha$} / \mbox{\rm H$\beta$}$ ratios, so we assume
a constant $\mbox{\rm H$\alpha$} / \mbox{\rm H$\beta$}$ ratio of 4.3
to correct the other regions through the formula:
$A_\lambda=\left(\frac{4.69}{\lambda}-1.45\right) \log\left(\frac{H
    \alpha/H \beta}{2.88}\right)$ \citep{osmer}.

We compared the brightness and colors of the knots to the evolutionary
synthesis models of \citet{bert}, with updatings (Girardi 1999,
private communication). We chose from their models those with a
Salpeter IMF and a heavy element abundance $Z=Z_\odot$.  In
Table~\ref{tab:phot_reg} we present the mean age of the knots obtained
by fitting our three colors to Girardi's models. Their masses
were derived from the V magnitude and assuming a distance of 85\,Mpc.
Within the errors all the knots, except for numbers 6 and 8, are
coeval to within our stated accuracy. The derived values, quoted in
Table~\ref{tab:phot_reg}, imply that we are dealing with star clusters
with masses\,$\leq$\,5$\times10^5$\,M$\odot$.

In Table~\ref{tab:flux} we present the \mbox{\rm H$\alpha$}
emission--line luminosities of all the knots in
Fig.~\ref{fig:zoomvel}, plus the nuclear emission of NGC\,6845\,A and
NGC\,6845\,B. As can be seen, knots 1, 2 and 3, located in the middle
of the tidal bridge, have \mbox{\rm H$\alpha$} luminosities comparable
to that of NGC\,3603, the most luminous \mbox{\rm H{\sc ii}} region
known to exist in our galaxy (L$_{\rm H_{\alpha}} = 1.1\times
10^{39}{\rm\,erg\,s}^{-1}$, \citealt{mtt}) or higher.  Regions 4 and
5, at the tip of the tidal bridge are even brighter, being similar to
30\,Doradus (L$_{\rm H_{\alpha}} = 6.3\times
10^{39}{\rm\,erg\,s}^{-1}$, \citealt{keh}). Giant \mbox{\rm H{\sc ii}}
regions near the ends of tidal bridges and tails were reported for
various systems, as in NGC\,4676 by \citet{stockton} and in NGC\,2623,
NGC\,6621/22, NGC\,3256 and NGC\,2535/36 by \citet{schweizer}.
These ionising clusters are usually more luminous than the ones along
the tidal tails.  Finally, \mbox{\rm H{\sc ii}} regions 7 and 8,
located at the extreme ends of the bar in NGC\,6845\,A, have
luminosities comparable to the ``Jumbo'' \mbox{\rm H{\sc ii}} region
of NGC\,3310 (L$_{\rm H_{\alpha}} =
1.4\times10^{40}{\rm\,erg\,s}^{-1},$ \citealt{past}), indicating a
strong star formation activity in these places, which may also have
been triggered by the collision \citep{kc}.

Like other interacting systems, this group is prominent in the
infrared. The IRAS fluxes (F$_{12\mbox{$\mu \rm\,m$}}=0.35$\,Jy,
F$_{25\mbox{$\mu \rm\,m$}}=0.32$\,Jy, F$_{60\mbox{$\mu
    \rm\,m$}}=2.78$\,Jy and F$_{100\mbox{$\mu \rm\,m$}}=6.66$\,Jy;
\citealt{rms}), lead to a group total luminosity of
$L_{IR}\,=7.9\times10^{10}{\rm\,L_\odot}$, locating this group at the
fainter end of Luminous IR galaxies according to the definition by
\citet{sm}.

%%%%%%%%%%%%%%%%
\subsection{\mbox{\rm H{\sc i}} gas distribution\label{subsec:h1dist}}
%%%%%%%%%%%%%%%%

In Fig.~\ref{fig:r_hi} we show the \mbox{\rm H{\sc i}} isodensity
contours superimposed on the R--band image. Despite the low spatial
resolution of the radio observations, it is possible to distinguish
two velocity components that differs by a few hundred
$\rm{km\,s}^{-1}$, a main component centered on NGC\,6845\,A and a
secondary one associated with NGC\,6845\,B. No \mbox{\rm H{\sc i}} was
detected at the velocities of galaxies C and D.

The main \mbox{\rm H{\sc i}} component presents a clear extension
towards the SW, bending South, following the stellar tidal tail. To
the East, it shows a protrusion that coincides with the bright optical
tidal bridge, with velocities compatible with this feature within the
observational errors.  The integrated \mbox{\rm H{\sc i}} flux of the
main component is $6.79{\rm{\,Jy\,km\,s}^{-1}}$, which translates to
an \mbox{\rm H{\sc i}} mass of $1.15\times10^{10}{\rm\,M_\odot}$.

The component associated with NGC\,6845\,B is virtually stationary in
velocity. The mean velocity is around 6700\,km\,s$^{-1}$, close to
that of NGC\,6845\,B. Spatially, it is slightly shifted eastward with
respect to NGC\,6845\,B.  The integrated \mbox{\rm H{\sc i}} flux is
$1.52{\rm{\,Jy\,km\,s}^{-1}}$, which translates to
$2.58\times10^{9}{\rm\,M_\odot}$.

The total integrated \mbox{\rm H{\sc i}} flux of both component is
$8.31{\rm{\,Jy\,km\,s}^{-1}}$, corresponding to an \mbox{\rm H{\sc i}}
mass of $1.41\times10^{10}{\rm\,M_\odot}$, or about 5 times more
\mbox{\rm H{\sc i}} than that contained in the Milky Way.

%%%%%%%%%%%%%%%%
\subsection{Kinematics\label{subsec:kin}}
%%%%%%%%%%%%%%%%

In Fig.~\ref{fig:slit} we show the slit positions along which we
obtained spectra. For ease of reference we indicated in
Fig.~\ref{fig:zoomvel} the radial velocity of the more relevant
objects. Radial velocities were obtained using a weighted mean of the
values measured for the \mbox{\rm H$\alpha$} , \mbox{\rm H$\beta$} and
{\mbox{[OIII]$_{\lambda\lambda 5007}$}} lines. A comparison with
Table~\ref{tab:gal} shows that the radial velocity values we measured
for NGC\,6845\,A and B are in agreement with previous results from
\citet{rosegraham} and \citeauthor{rc3}. There is, however, a
discrepancy between the values given by the cited references for
radial velocities of components C and D.  Apparently one of this
sources made a mistake in the galaxies identification.

%%%%%%%%%%%%%%%%
\subsubsection{NGC\,6845\,A: Optical rotation curve\label{subsubsec:optkin}}
%%%%%%%%%%%%%%%%

The rotation curve of NGC\,6845\,A, presented in
Figure~\ref{fig:rota}, was obtained from slit position 4 (see
Table~\ref{tab:spec} and Fig.~\ref{fig:slit}), along the apparent
major axis of the galaxy. Line of sight velocity measurements were
transformed to circular velocities adopting an inclination angle of
$57\degr (\pm 3\degr)$, derived by fitting ellipses to the optical
isophotes of the B image (using the appropriated procedure in the {\sc
  STSDAS--IRAF} package). Since the isophotes in the inner part of
NGC\,6845\,A are influenced by the bar profile, and those more
external are tidally distorted, we adopted the ellipticity of
isophotes at intermediate radii as the best indicator for the
inclination. There is a marked difference between the NE and the SW
sides of the rotation curve for r$\geq\,$6\,kpc (see
Fig.~\ref{fig:rota}): to the NE it shows a strong bump at about
6\,kpc, whereas the SW half presents a plateau at about the same
distance. A qualitative comparison with numerical simulations 
\citep[e.g.][]{mb,barton} suggests that the difference is
more likely due to tidal rather than intrinsic effects, which perturb
more strongly the galactic disk periphery.

The average rotation curve (Fig.~\ref{fig:rota}) was fitted with
\citeauthor{satoh}'s (1980) density law:

$$
V_c^2(R,Z)\,=\,GMR^2\,[\,R^2+Z^2+a(a+2\,\sqrt{Z^2+b^2})]^{-3/2}
$$

\noindent where $V_c(R,Z)$ is the circular velocity at position
$(R,Z)$, M is the total mass and (a,b) are shape parameters with the
dimension of length. Following Satoh, a ratio of $b/a\approx0.1$
corresponds to a disk dominated distribution, $b/a\approx1$ to a disk
plus a well developed bulge  \citep[Sa--SO, see][]{satoh} and
$b/a\approx10$ to a near spherical distribution. $\chi^2$ fitting was
performed by the downhill simplex minimization algorithm (``amoeba'')
with the IRAF NFIT1D task. The parameters obtained for the fit were
$a=6.5 \pm 0.4$ and $b/a=0.3 \pm 0.1$, which indicates a dominant disk
with a small to intermediate bulge, in agreement with \citeauthor{rc3}
classification The fit is shown in Fig.~\ref{fig:rota}, where the full
line represents the best fit and the dashed lines give its lower and
upper limits within the observational errors, which in turn furnishes
$a$ and $b/a$ errors. The derived model total mass is $M_{Satoh}=4.4
(\pm 1.2) \times10^{11}{\rm\,M_\odot}$. The Keplerian mass up to the
last point on the rotation curve is $3.2\times10^{11}{\rm\,M_\odot}$,
broadly in agreement with the more sophisticated fit based on
\citeauthor{satoh}'s model.

NGC\,6845\,A mass is comparable to that of Andromeda. Andromeda's
brightness is $V=3.44$\,mag at a distance of 0.7\,Mpc, corresponding
to 13.5\,mag at the distance of NGC\,6845\,A, and indicating that both
galaxies are comparable in their M/L ratio. Nevertheless, the size of
NGC\,6845\,A is only half that of Andromeda. This figure would be
difficult to change since in order to make both galaxies comparable in
size, a Hubble constant smaller than $35{\rm{\,km\,s^{-1}\,Mpc^{-1}}}$
would be required. It is also difficult to invoke an error in the
derived mass for NGC\,6845\,A due to an error in the disk inclination
since, even by assuming an edge-on position for this galaxy, we could
not reduce the derived mass by more than a factor of 1.5. \citet{gr}
pointed already out that NGC\,6845\,A should have a large mass because
of its uncommonly large maximum velocity. We note that their rotation
curve extends only up to $\sim 6kpc$.

%%%%%%%%%%%%%%%%
\subsubsection{Tidal bridge kinematics\label{subsubsec:arm}}
%%%%%%%%%%%%%%%%
It is interesting to note that the tidal bridge of NGC\,6845\,A shows
a decreasing trend in radial velocity (see Fig~\ref{fig:zoomvel} and
Tab~\ref{tab:flux}) starting from its base, where it is connected to
the NGC\,6845\,A disk near region 8 (at 6660${\rm{\,km\,s}^{-1}}$),
and passing through \mbox{\rm H{\sc ii}} regions 3, 2 and 1, located
more or less in the middle of the tidal bridge (at
6455${\rm{\,km\,s}^{-1}}$, 6361${\rm{\,km\,s}^{-1}}$ and
6272${\rm{\,km\,s}^{-1}}$, respectively). Regions 4 and 5, which are
slightly shifted to the north with respect to the prolongation of the
bridge, surprisingly present radial velocities that agree with this
trend (6314${\rm{\,km\,s}^{-1}}$ and 6201${\rm{\,km\,s}^{-1}}$,
respectively), strongly suggesting that they are linked to the tidal
bridge and not to NGC\,6845\,B, despite the fact that they are seen
projected on top of this object.  We also detected \mbox{\rm H{\sc i}}
on the tidal bridge, with a mean velocity of
6350${\rm{\,km\,s}^{-1}}$, in agreement with the optical results (see
next Section).

Supposing that the tidal bridge tip follows a parabolic orbit in the
plane of NGC\,6845\,A disk and its present true velocity is the
observed radial velocity of the \mbox{\rm H{\sc ii}} regions 4 and 5,
we obtain a collision age of $\sim$\,150---300\,Myr. Since \mbox{\rm
  H{\sc ii}} regions cannot be older than 10\,Myr (see
Sec.~\ref{subsec:knots}), this implies that star formation activity
occurs inside the tidal bridge, started well after the beginning of
the bridge formation.

A possible scenario for the bridge late star formation come out from
\citet{bh} simulations. These authors found that self--gravitating
star clumps forms along tidal tails after a collision. The gas clouds
torn apart from the galactic disk simultaneously with the stars may
fall into these clumps, reaching the necessary density in order to
form stars. Cloud--cloud collisions inside the tidal bridge may be an
alternative mechanism to explain the star formation inside the bridge.

%%%%%%%%%%%%%%%%
\subsubsection{\mbox{\rm H{\sc i}} velocity field of NGC\,6845\,A\label{subsubsec:h1vel}}
%%%%%%%%%%%%%%%%

The isovelocity map of the \mbox{\rm H{\sc i}} component associated
with NGC\,6845\,A is shown in Fig.~\ref{fig:r_vel}, superimposed on
the R--band image.
 
In Fig.~\ref{fig:prof}, we show the \mbox{\rm H{\sc i}}
position--velocity diagram, obtained by cutting the data cube along
the major axis of NGC\,6845\,A (line of the optical spectrum slit
number 4 in Figure~\ref{fig:slit}).  In this figure the center of
NGC\,6845\,A is at $\delta=-47^\circ12'30''$ and $6399{\rm
  \,km\,s^{-1}}$.  To the South of the galaxy center, the diagram
shows a behavior similar to that of the optical rotation curve
(Fig.~\ref{fig:rota}), so we conclude that in this part the \mbox{\rm
  H{\sc i}} kinematics is mainly rotational.  To the North, the lower
part of this diagram shows a plateau at $6350{\rm \,km\,s^{-1}}$,
coincident with the tidal bridge velocity. Supposing that the
\mbox{\rm H{\sc i}} is in dynamical equilibrium, we derive from its
kinematics (Fig.~\ref{fig:r_vel}) a mass of
1.4$\times10^{12}M_{\odot}$, about three times higher than the mass
derived from the optical data. We attribute this difference to the
fact that the \mbox{\rm H{\sc i}} can be followed out to larger
distances, especially to the SW (the upper part of the
position--velocity diagram of Fig.~\ref{fig:prof}), and hence will
trace the dark matter associated with the halo of NGC\,6845\,A.
Nevertheless, we have to keep in mind that the \mbox{\rm H{\sc i}}
kinematics may suffer non--rotational effects. The \mbox{\rm H{\sc i}}
map low spatial resolution does not allow to see the classical
butterfly diagram, characteristic of 2-D rotational maps. As can be
seen in Figure~\ref{fig:r_vel}, the isovelocity lines are roughly
oriented along the beam major axis.

%%%%%%%%%%%%%%%%
\subsubsection{The group kinematical mass\label{subsubsec:kmass}}
%%%%%%%%%%%%%%%%

Following \citet{binney}, equation 10--21), the radial velocity
difference among the system members, their absolute luminosities
(Tab.~\ref{tab:phot_gal}) and projected positions, allow us to derive
the group M/L ratio. For the group member velocities we adopt the mean
value of the available data (\citealt{rosegraham}; \citeauthor{rc3} and
ours, see Tab.~\ref{tab:phot_gal}).

An important source of error in determining M/L is the high
inclination of the C component and, to a less extent, that of the A
component, the group most massive galaxies.  The inclination of each
group member was obtained as quoted in Sec.~\ref{subsubsec:optkin} for
NGC\,6845\,A. \citeauthor{holmberg}'s (1958) correction was applied to
the measured isophotal axis ratio $b/a$ by the formula
cos$^2(i)=((b/a)^2-q_0^2)/(1-q_0^2)$ in order to obtain the galaxies'
inclinations (see Tab.~\ref{tab:phot_gal}), where $q_0=0.2$
\citep{kw}. This correction is specially important for NGC\,6845\,C.
We obtained for this galaxy $b/a=0.22\pm0.05$, measured at the
isophotal level 25 mag/arcsec$^2$ . Galactic extinction was corrected
with the standard absorption law \citep{buh}. In order to obtain the
galaxies face--on magnitudes we corrected internal absorptions by the
empirical (\citeauthor{rc3}) formula $A(i)=A(0)+\alpha \log \sec (i)$,
which was evaluated following two different prescriptions of the
values of $A(0)$ and $\alpha$. The first one is from \citet{rftb},
where $A(0)=0$ and $\alpha=1.9$, leading to (M/L)$_{1}=43.5 \pm 2.0$
(solar units), with the uncertainties estimated from the errors in
$\alpha$ $(\pm 0.16$, \citet{kw} and $i$ (Tab.~\ref{tab:phot_gal}).
The second prescription is that of \citeauthor{rc3}, where $A(0)=0.12$
and $\alpha=0.8$ for NGC\,6845\,A and $A(0)=0.1$ and $\alpha=0.7$ for
galaxies B, C and D. In this case we obtained (M/L)$_{2}=66.4 \pm
2.0$. \citet{rosegraham} estimated M/L in the range 28---49,
also depending on the source for Galactic and internal extinction
corrections. Our (M/L)$_{1}$ is well inside their range of values,
while (M/L)$_{2}$ is higher than their highest M/L ratio.

Although large, the difference between (M/L)$_{1}$ and (M/L)$_{2}$
points to a very similar total mass of the system ($1.0\pm 0.2\times
10^{13}\,M_\odot$ and $8.7\pm 0.2\times 10^{12}\,M_\odot$
respectively). The luminous mass of the system cannot account for more
than $\approx\,1.6\times 10^{12}\,M_\odot$ (scaling the mass of the
galaxies by their luminosities with respect to the mass and luminosity
of NGC\,6845\,A), indicating that the group's dark matter amounts to
approximately 5 times higher than its luminous mass. The whole system
contains almost 9 times more dark matter than what was thought to be
associated with the halo of NGC\,6845\,A, as derived in the previous
Section from \mbox{\rm H{\sc i}} kinematics.

%%%%%%%%%%%%%%%%
\section{Scenarios for the collision\label{sec:dynamics}}
%%%%%%%%%%%%%%%%

Strong, well developed tidal bridges and tails are characteristic of
galaxy encounters in which the disk of the perturbed galaxy and the
orbit of the perturber rotate in the same sense, i.e., a prograde
orbit, and the angle between the main galaxy disk and the orbital
plane is small \citep[e.g.][]{tt,nw,hkbb}.  In prograde encounters,
tidal arms curve in a sense opposite to the disk rotation: they are
trailing.  Consequently, we can infer that NGC\,6845\,A disk rotates
counterclockwise (Sec.~\ref{subsubsec:optkin}) and its SE side is
preceding, indicating that the tidal bridge side is the near one.

\citet{gr} already noted as curious that due to the relative
position and radial velocities of galaxies A and B, any simple
geometry would indicate that the B orbit is retrograde. However, as
mentioned by these authors, the case might be similar to that of
M\,51/NGC\,5195, for which~\citet{tt} have found it possible to
obtain a direct sense of rotation if the orbital eccentricity of
NGC\,5195 is large and if the orbit is highly inclined to the plane of
M\,51. From this point of view, galaxy C would be a good candidate for
the strong interaction. Nevertheless, this galaxy presents itself as a
symmetric disk up to faint brightness levels. Modern numerical
simulations of similar disk encounters \citep[e.g.][]{barnes92},
shows that a collision between two large disks, like those of
NGC\,6845\,A and C, close enough as to produce NGC\,6845\,A tidal
bridge, would necessarily strongly distort NGC\,6845\,C disk as well.

In order to understand the dynamical evolution of this group, two of
us \citep[][, in preparation]{simul},  are developing extensive
numerical simulations with the TreeSPH code \citep{hk} in a CRAY
T94 computer (CESUP--UFRGS). From this work we advance some
preliminary results. In Figs.~\ref{fig:simulac} and~\ref{fig:simulab},
we shows simulations of A with C and A with B collisions.  Galaxy
models were constructed following the general prescription of
\citet{kd}.  Galaxy A model is such that its rotation curve
presents a circular velocity of 300\,km\,s$^{-1}$ at 6\,kpc, and is
approximately flat up to the cutoff radius (20\,kpc). 10\% of the disk
mass is composed of gas particles, whose density profile is more
extended than that of the stellar disk. For galaxy C we adopted the
same model of A, but without gas, since this galaxy do not present
\mbox{\rm H{\sc i}} or emission lines in its optical spectrum. A and C
galaxy models consist of 6000 particles with a mass of 3.5 $\times
10^{11}\,M_\odot$ per galaxy.  Galaxy B is represented by a point mass
of 1/4 of model A mass.

Fig.~\ref{fig:simulac} shows the evolution of the encounter of models
A and C, which approach along a parabolic orbit with a perigalactic
distance $q=12$\,kpc. The first row shows the time evolution of the
stellar component, while the second row shows the same for the gas
component. For easy of comparison between simulation and observations
we used a coordinate system in which the z--axis corresponds to the
line--of--sight (with positive values towards the observer), the
x--axis is the E-W direction (positive to the West) and the x--axis is
the N-S direction (positive to the North). In this system, the orbit
of C around A has its pericenter in the direction of the vector
$(-3,-10,10)$, and the orbital plane is defined by the orbital spin
vector, parallel to $(10,3,6)$. We identify the third frame of
Fig.~\ref{fig:simulab} (104\,Myr after the perigalacticum) as the more
similar to the system present stage. A striking result in favor of
this scenario is the presence of a stellar condensation similar to
that observed to the right of galaxy C NW extreme.  C model SE extreme
bend to the E, which is also observed to the faintest brightness
levels in galaxy C (Fig.~\ref{fig:rimag}).  The simulation also
clearly shows that this collision produces a strong tidal feature
similar to the NGC\,6845\,A tidal bridge with a condensation near its
extreme that resembles \mbox{\rm H{\sc ii}} regions 4 and 5, but
nothing as large as galaxy B is produced out of this encounter, so the
position of galaxy B at the tidal bridge extreme would be a projection
within this scenario.  Against this scenario the simulation shows that
about 10\% of the gas originally belonging to galaxy A would be
captured by galaxy C, which is neither observed as \mbox{\rm H{\sc
    ii}} nor as \mbox{\rm H{\sc i}}. The simulation also shows that
stars stripped out from A should form a stellar feature perpendicular
to C disk, also not detected in the optical images, which present a
high degree of symmetry to intermediate brightness levels.

Fig.~\ref{fig:simulab} shows the encounter of model A with the point
mass B, under a prograde hyperbolic orbit with eccentricity $e=2$ and
perigalactic distance $q=8$\,kpc. The pericenter is in the direction
$(20,-8,-10)$ and its orbital spin vector is parallel to $(8,20,0)$.
The third frame (145\,Myr after the perigalacticum, shown in the
second frame) shows the best match of the model to the system present
stage. In this frame galaxy B is receding with respect to A at
$300\,{\rm km\,s}^{-1}$. The more striking feature of this simulation
is the morphology of the tidal bridge, which is straight and narrow
and shows condensations of gas and stars at its base and middle part,
resembling the observed tidal bridge morphology. Also the tidal tail
is diffuse, as in the real situation. Against this scenario the
simulation shows that the tidal tail points in a direction different
from the observed one and does not bifurcate.

In spite of both scenarios pros and cons, the collision of A and B
seems to be more appropriate to explain the tidal bridge structure.
Although some features associated to galaxy C can be regarded as
tidally produced, it cannot be ignored the lack of other important
features that C should show if it were the main perturber. Perhaps a
three--body interaction, with B producing the tidal bridge would be
more suitable to explain the richness of features observed in this
group.

%%%%%%%%%%%%%%%%
\section{Concluding remarks\label{sec:conclu}}
%%%%%%%%%%%%%%%%
The system NGC\,6845 is a compact group in which NGC\,6845\,A presents
a strong tidal bridge, signature of a close collision.

We present the first atomic Hydrogen observation of NGC\,6845,
obtaining an \mbox{\rm H{\sc i}} content 5 times higher than in the
case of the Milky Way. The \mbox{\rm H{\sc i}} distribution can be
separated into two components, one related to NGC\,6845\,A, sporting
well developed tidal features and the second one, with velocity
similar to that of NGC\,6845\,B.  Components C and D do not present
\mbox{\rm H{\sc i}}, up to our detection limit.  This fact set
constrains to galaxy C being the main perturber of A, since, as shown
by numerical simulations, in these case gas originally belonging to A
should be transferred to C.

NGC\,6845\,A optical rotation curve is asymmetric, with a uncommonly
high maximum circular velocity, especially to the SW side. Satoh model
fitting to the rotation curve indicates the presence of a dominant
disk and a small to intermediate bulge, in agreement with the
morphological classification of this galaxy. The derived mass is
M$_{Satoh}=4.4(\pm1.2)\times 10^{11}{\rm\,M_\odot}$. The (B-I) color
map indicate the presence of dust in front of the SW side of the bar,
pointing to a trailing bar.

The (B-I) color map also shows that the bridge is bluer than
NGC\,6845\,A inner disk.  \mbox{\rm H{\sc ii}} regions brighter than
those of the Antennae and NGC\,7252 tidal tails have formed at the tip
and at the base of the tidal bridge.  Three \mbox{\rm H{\sc ii}}
regions midway along the tidal bridge are brighter than NGC\,3603, the
brightest one known in our galaxy.  The kinematics of \mbox{\rm H{\sc
    ii}} regions 4 and 5 indicates that they are more likely connected
to the tidal bridge than to NGC\,6845\,B, as their projected position
may suggest. They present luminosities similar to that of 30\,Doradus.
A simple model for the kinematics of the tip of the bridge as well as
numerical simulations indicate that star formation is occurring along
this structure one hundred million years or more after the perturber
closest approach.
 
The group kinematics calls for dark matter within the system at about
5 times its luminous mass. The group amount of dark matter would be 9
times higher than the halo mass of the NGC\,6845\,A, provided that the
\mbox{\rm H{\sc i}} is in dynamical equilibrium inside the galaxy;
this in turn would be three times higher than the galaxy kinematical
mass derived from the optical rotation curve.

Since abundant signs of interactions are present in galaxy B, and to a
less extent in C, we appealed to N-body simulations in order to
disentangle which of the galaxies, B or C, would be the A main
perturber. Both scenarios for the tidal disruption of galaxy A present
pros and cons, but the bridge, which is the most conspicuous tidal
feature of this galaxy, seems to be better reproduced by the
simulation of the encounter of galaxies A and B. The fact that neither
\mbox{\rm H{\sc i}} nor \mbox{\rm H{\sc ii}}, that should have been
stripped out from galaxy A, are detected in galaxy C reinforce this
scenario.  Perhaps a three body interaction would be a more
appropriate scenario to explain the group dynamical evolution, but it
is difficult to model due to the large number of degrees of freedom
involved such a collision.

%%%%%%%%%%%%%%%%%%%%%%%%%
\acknowledgements We are grateful to Dr. Lars Hernquist, who kindly
made available the TreeSPH code, to Mr. Ruben Diaz, who carried out
the observations at Bosque Alegre and to Dr. Leo Girardi, who kindly
provided his evolutionary synthesis models. The authors greatefully
acknowledge the NRAO for allocating observing time and for support. We
also thank the staff of CASLEO and Bosque Alegre. Numerical
calculations were done at the CESUP-UFRGS national supercomputing
center. This work was supported in part by the Brazilian institutions
CNPq and FINEP.

%%%%%%%%%%%%%%%%%%%%%%%%%

%%%%%%%%%%%%%%%%%%%%%%%%%

%%%%%%%% Tables
\clearpage

\begin{deluxetable}{l c c c c l}
  \tablewidth{0pt}
  \tablecaption{Basic data on the four components A, B, C, and D of
    the NGC\,6845 quartet (see Fig.~\ref{fig:rimag}).\label{tab:gal}}
  \tablehead{
    \colhead{ } &
    \colhead{A} &
    \colhead{B} &
    \colhead{C} &
    \colhead{D} &
    \colhead{References} 
    }
  \startdata
  $\alpha$(1950)    & $19^h57^m22.0^s$  & $19^h57^m29.6^s$  &
  $19^h57^m20.3^s$  & $19^h57^m17.0^s$  & \citeauthor{rc3} \\
  $\delta$(1950)    & $-47^\circ12'30''$& $-47^\circ11'53''$&
  $-47^\circ13'20''$& $-47^\circ13'56''$& \citeauthor{rc3} \\
  l                 & 352.24            & 352.26             &
  352.23            & 352.21            & \citeauthor{rc3} \\
  b                 & -30.99            &  -31.02            &
  -30.99            & -30.98            & \citeauthor{rc3} \\
  $m_B$ (mag)       & $13.92$    & $15.32$    &
  $14.94$     & $16.39$    & \citet{rosegraham} \\
  $B_T$ (mag)       & $13.65\pm0.15$    & $14.86\pm0.15$    &
  $16.3\pm0.30$     & $15.50\pm0.20$    & \citeauthor{rc3} \\
  $m_B$ (mag)       & $13.84\pm0.02$    & $15.19\pm0.02$    &
  $14.35\pm0.02$     & $15.95\pm0.02$    & This paper (see Tab.~\ref{tab:phot_gal}) \\
  $V_{sys} ({\rm{\,km\,s}^{-1}})$& $6410$ & $6776.5$
  & $6755$ & $7070$           &  \citet{rosegraham} \\
  $V_{sys} ({\rm{\,km\,s}^{-1}})$& $6356\pm18$ & $6753\pm18$
  & $7070\pm54$ & $6816\pm44$           &  \citeauthor{rc3} \\
  $V_{sys} ({\rm{\,km\,s}^{-1}})$& $6399\pm20$ & $6746\pm20$
  & \nodata         & \nodata           & This paper (see Tab.~\ref{tab:flux}) 
  \tablecomments{$V_{sys}$ is heliocentric. Units of right ascension
    are hours, minutes and seconds, and units of declination are
    degrees, arcminutes and arcseconds.}
  \enddata 
\end{deluxetable}

\begin{deluxetable}{c c c c}
  \tablewidth{0pt}
  \tablecaption{Log of broad-band CCD--imaging observations. \label{tab:ima}}
  \tablehead{
    \colhead{Filter} &
    \colhead{Date}   & 
    \colhead{Number} &
    \colhead{Exp.\,time\tablenotemark{a}}\\
    \colhead{ }      &
    \colhead{ }      &
    \colhead{of frames} &
    \colhead{(sec.)}  
    }
  \startdata
  B      & Sep 06, 1997 & 7         & 2100     \\
  V      & Sep 06, 1997 & 6         & 1800     \\
  R      & Sep 06, 1997 & 4         & 720      \\
  I      & Sep 06, 1997 & 4         & 360      
\tablenotetext{a}{Total integration time.}  \enddata
\end{deluxetable}

\begin{deluxetable}{c c c c c c r c c}
  \tablewidth{0pt}
  \tablecaption{Log of spectroscopic observations. \label{tab:spec}}
  \small
  \tablehead{
    \colhead{Pos}       &
    \colhead{Date}      & 
    \colhead{Spectrograph} & 
    \colhead{Grating}   & 
    \colhead{Spectral}  &
    \colhead{Resolution}   &
    \colhead{Slit size} &
    \colhead{Number of} & 
    \colhead{Total\,exp.} \\
    \colhead{}          &
    \colhead{}          &
    \colhead{}          &
    \colhead{(l\,mm$^{-1}$)} &
    \colhead{range (\AA)} &
    \colhead{(\AA)}     &  
    \colhead{(arcsec)}  &
    \colhead{frames}    & 
    \colhead{time (sec.)}
    }
  \startdata
  1    & Jul 08 1994 & B \& C       &300      & 4100-6900 & 8 &
  4.5$\times$170 &  3      & 4800     \\
  2    & Jul 08 1994 & B \& C       &300      & 4100-6900 & 8 &
  6$\times$170   &  3      & 10800    \\
  2    & Sep 08 1997 & REOSC        &600      & 5870-7460 & 5 &
  3.6$\times$120 &  1      & 3600     \\
  2$^*$& Dec 18 1997 & Multi-function &1200 &6130-6850& 2 &
  1$\times$130   &  1      & 3600     \\
  3    & Jul 09 1994 & B \& C       &300      & 4100-6900 & 10&
  2$\times$170   &  3      & 3200     \\
  4    & Jul 10 1994 & REOSC        &300      & 3700-7150 & 8 &
  2$\times$120   &  2      & 1800     \\
  5    & Jul 10 1994 & REOSC        &300      & 3700-7150 & 8 &
  6$\times$120   &  2      & 2700     \\
  6    & Sep 08 1997 & REOSC        &600      & 5870-7460 & 6 &
  2.3$\times$120 &  4      & 9300     \\
  7    & Sep 08 1997 & REOSC        &600      & 5870-7460 & 6 &
  2.3$\times$120 &  1      & 3600     
  \tablecomments{Slit positions
    given in the first column are those marked in
    Figure~\ref{fig:slit}.  The spectrum marked with an asterisk was
    observed at the Bosque Alegre Astrophysical Station; all other
    observations were performed at CASLEO.}
  \enddata
\end{deluxetable}

\begin{deluxetable}{c c c c c c c c c c c c c}
  \tablewidth{0pt}
  \tablecaption{Integrated B magnitudes and colors in Johnson's system
    of the group members. All galaxies were measured through a
    64\farcs8 aperture. \label{tab:phot_gal}}
  \scriptsize
  \tablehead{
    \colhead{Gal.}     &
    \colhead{B}     & 
    \colhead{(B-V)} & 
    \colhead{(V-R)} & 
    \colhead{(V-I)} &
    \colhead{ $b/a$ } &
    \colhead{$\Delta m_B$} &
    \colhead{$\Delta m_i^{(1)}$} &
    \colhead{$\Delta m_i^{(2)}$} &
    \colhead{M$_B^{(1)}$} &
    \colhead{M$_B^{(2)}$} &
    \colhead{L$_B^{(1)}$} &
    \colhead{L$_B^{(2)}$} \\
    \colhead{(1)}     &
    \colhead{(2)}     & 
    \colhead{(3)} & 
    \colhead{(4)} & 
    \colhead{(5)} &
    \colhead{(6)} &
    \colhead{(7)} &
    \colhead{(8)} &
    \colhead{(9)} &
    \colhead{(10)} &
    \colhead{(11)} &
    \colhead{(12)} &
    \colhead{(13)} 
    }
  \startdata

  A         & 13.84     & 0.64        & 0.23      &
  0.20      & 0.54      & 0.16 & 0.50 & 0.32 & -21.56 & -21.38 & 3.2 
            & 2.7              \\  

            & $\pm$0.02 & $\pm$0.03   & $\pm$0.07 &
  $\pm$0.14 & $\pm$0.06 &      &      &      &        &        & $\times
  10^{40}$  & $\times 10^{40}$ \\

  B         & 15.19     & 0.64        & 0.26          &
  0.29      &  1        & 0.16 & 0.00 & 0.10 & -19.71 & -19.81 & 5.9
            & 6.5              \\  

            & $\pm$0.02 &  $\pm$0.03  &     $\pm$0.10 &
  $\pm$0.22 & $\pm$0.08 &      &      &      &        &        & $\times
  10^{39}$  & $\times 10^{39}$ \\  

  C         & 14.35     & 0.81           & 0.35          &
  0.16      & 0.22      & 0.16 & 1.86 & 0.82 & -22.41 & -21.37 & 7.1
            & 2.7              \\  

            & $\pm$0.02 &     $\pm$0.04  &     $\pm$0.10 &
  $\pm$0.17 & $\pm$0.05 &      &      &      &        &        & $\times
  10^{40}$  & $ \times 10^{40}$ \\  

  D         & 15.95     & 0.83           & 0.35          &
  0.11      & 0.33      & 0.16 & 1.06 & 0.50 & -20.01 & -19.45 & 7.8
            & 4.6               \\  

            &      $\pm$0.02  &     $\pm$0.04  &     $\pm$0.08 &
  $\pm$0.18 & $\pm$0.03 &      &      &      &        &        & $\times
  10^{39}$  & $\times 10^{39}$  \\  
  
  \enddata 
  \tablecomments{Col.\,(1): NGC\,6845 group individual galaxy
    names. Col.\,(2): apparent blue magnitudes. Cols.\,(3), (4) and
    (5): colors. Col.\,(6): corrected inclination angle (see
    Sec.~\ref{subsubsec:kmass}). Col.\,(7):corrections to apparent B
    magnitude for Galactic extinction \citep{buh}. Cols.\,(8) and (9):
    internal extinction corrections to face--on magnitudes, following
    \citet{rftb} and \citeauthor{rc3}, respectively (see
    Sec.~\ref{subsubsec:kmass}). Cols.\,(10) and (11): absolute blue
    magnitudes with corrections for internal absorption given by
    $\Delta m_i^{(1)}$ and $\Delta m_i^{(2)}$, respectively. We used
    the mean Galactocentric velocity of the group members as distance
    indicator. Cols.\,(12) and (13): absolute blue luminosity $({\rm
      erg\,s^{-1}})$ as taken from cols. (10) and (11), respectively
    (H$_0=75{\rm{\,km\,s^{-1}\,Mpc^{-1}}}$).}
\end{deluxetable}

\begin{deluxetable}{c c c r r c c}
  \tablewidth{0pt} 
    \tablecaption{Integrated V magnitudes and colors for the
    condensations described in the text (see also
    Figure~\ref{fig:zoomvel}). \label{tab:phot_reg}}
  \tablehead{
    \colhead{Region} & 
    \colhead{V}      & 
    \colhead{(B-V)}  & 
    \colhead{(V-R)}  & 
    \colhead{(V-I)}  & 
    \colhead{Age $(10^6yr)$} & 
    \colhead{Mass $(10^4{\rm\,M_\odot})$} 
    }
  \startdata
  1      & 16.38$\pm$0.02  & 0.17$\pm$0.04  &  0.06$\pm$0.09 & 
  0.27$\pm$0.11 & $6.9\pm1.0$ & $21\pm5$ \\
  2      & 16.43$\pm$0.04  & 0.15$\pm$0.08  &  -0.02$\pm$0.12 & 
  0.12$\pm$0.22 & $6.3\pm1.0$ & $20\pm5$ \\
  3      & 16.45$\pm$0.01  & 0.20$\pm$0.02  &  0.00$\pm$0.12 & 
  0.15$\pm$0.13 & $6.5\pm1.0$ & $19\pm5$ \\
  4      & 16.03$\pm$0.02  & 0.28$\pm$0.03  & -0.05$\pm$0.08 &
  -0.30$\pm$0.16 & $6.6\pm1.0$ & $28\pm5$ \\
  5      & 16.18$\pm$0.02  & 0.26$\pm$0.03  &  0.05$\pm$0.06 &
bn  -0.28$\pm$0.12 & $6.6\pm1.0$ & $25\pm5$ \\
  6      & 18.52$\pm$0.06  & 0.27$\pm$0.09  &  0.11$\pm$0.19 & 
  0.50$\pm$0.43 & $8.0\pm1.0$ & $1.3\pm5$ \\
  7      & 14.57$\pm$0.01  & 0.26$\pm$0.02  & -0.13$\pm$0.05 &
  -0.48$\pm$0.09 & $5.3\pm1.0$ & $51\pm5$ \\
  8      & 14.81$\pm$0.01  & -0.37$\pm$0.02 & -0.10$\pm$0.06 &
  -0.40$\pm$0.09 & $<3.0\pm1.0$& $40\pm5$
  \tablecomments{All regions were measured through a
    $4''$ aperture. Magnitudes were corrected for Galactic extinction
    with the standard absorption law, adopting $A_B=0.16$ \citep{buh}
    and for internal absorption using $\mbox{\rm H$\alpha$} /
    \mbox{\rm H$\beta$} = 4.3$ (see Sec.~\ref{subsec:knots}).}
  \enddata
\end{deluxetable}

\begin{deluxetable}{c c c c c c}
  \tablewidth{0pt} 
  \tablecaption{\mbox{\rm H$\alpha$} luminosities,
    \mbox{\rm H$\alpha$} /\mbox{\rm H$\beta$} ratios, slit dimensions
    and positions, and Radial velocities for the condensations marked
    in Fig.~\ref{fig:zoomvel}.  \label{tab:flux}} 
  \tablehead{
    \colhead{Region} & 
    \colhead{L$_{\rm H_{\alpha}}$} & 
    \colhead{\mbox{\rm H$\alpha$} / \mbox{\rm H$\beta$} } & 
    \colhead{Slit dimension} & 
    \colhead{Slit pos.}  &
    \colhead{Vel.}\\
    \colhead{} & 
    \colhead{$({\rm\,erg\,s}^{-1})$} & 
    \colhead{} & 
    \colhead{(arcsec)} &
    \colhead{} & 
    \colhead{(km\,s$^{-1}$)}\\
    \colhead{(1)} & 
    \colhead{(2)} & 
    \colhead{(3)} & 
    \colhead{(4)} &
    \colhead{(5)} & 
    \colhead{(6)} } 
  \startdata
  1  & $2.3\times10^{39}$ & \nodata  & $6(1)\times6$ & 2   & 6272 \\
  2  & $3.0\times10^{39}$ & \nodata  & $6(1)\times6$ & 2   & 6361 \\
  3  & $1.8\times10^{39}$ & \nodata  & $6(1)\times6$ & 2   & 6455 \\
  4  & $9.1\times10^{39}$ & 4.2      & $2.3\times6$ &7;3;5& 6314 \\
  5  & $6.5\times10^{39}$ & \nodata  & $2.3\times6$ & 6   & 6201 \\
  6  & \nodata            & \nodata  & \nodata         & \nodata & \nodata \\
  7  & $2.2\times10^{40}$ & 4.1      & $2\times6.2$ & 4   & 6113 \\
  8  & $8.7\times10^{39}$ & 4.6      & $2\times6$      & 4   & 6660 \\
  NGC\,6845\,A & $7.7\times10^{39}$  & \nodata & $2\times10$ & 4   & 6399 \\
  NGC\,6845\,B & $1.1\times10^{40}$  & \nodata & $2.3\times8$ & 6;7 &6746 \\
  \enddata 
  \tablecomments{Col.\,(1): region label as in
    Table~\ref{tab:phot_reg} and Fig.~\ref{fig:zoomvel}. Cols.\,(2)
    and (3): \mbox{\rm H$\alpha$} luminosities and \mbox{\rm
      H$\alpha$} / \mbox{\rm H$\beta$} ratios, respectively.
    Col.\,(4): slit dimensions. The numbers in parentheses on the slit
    dimension of regions 1, 2 and 3 correspond to the slit width of
    the spectrum used to measure velocities, while the values outside
    the parentheses is for the spectra used to measure flux.
    Col.\,(5): slit position, as in Table~\ref{tab:spec}. In region 4,
    three slit positions were used: position 7 to measure \mbox{\rm
      H$\alpha$} flux and radial velocity and positions 3 and 5 to
    measure \mbox{\rm H$\alpha$}-to-\mbox{\rm H$\beta$} ratio.
    Col.\,(6): heliocentric radial velocity of the region.}
\end{deluxetable}

%%%%  Figures %%%%%%%%%%%%%%%%%%%%%%%
\clearpage

\figcaption[RODRIGUES.fig01.ps]{A 600\,s R--band image of the group of
  galaxies NGC\,6845 or Klemola\,30. The contrast is chosen
  so as to emphasize low surface brightness levels. A very good photo
  of this group can be seen in \cite{rose} (1979). \label{fig:rimag}}

\figcaption[RODRIGUES.fig02.ps]{(B-I) color map. 10 grey levels are
  shown, ranging from 0.5 (white) to 3.5 mag\,pix$^{-1}$ (black).
  \label{fig:colorBI}}
  
\figcaption[RODRIGUES.fig03a.ps,RODRIGUES.fig03b.ps]{The same as
  Fig.~\ref{fig:rimag}, detatching bright condensations. The circles
  show the positions of the \mbox{\rm H{\sc ii}} regions whose photometric and
  spectroscopic measurements are presented in
  Tables~\ref{tab:phot_reg} and~\ref{tab:flux}. Left hand panel:
  radial velocities of the galaxies A, B, C and D, as well as of some
  selected positions. Right hand panel: detail of the tidal bridge,
  showing \mbox{\rm H{\sc ii}} regions 1, 2 and 3. \label{fig:zoomvel}}

%%% Figuras 4 e 5 foram trocadas pelo editor
\figcaption[RODRIGUES.fig05.ps]{Composite image showing our optical
  spectra towards seven of the eight \mbox{\rm H{\sc ii}} regions indicated in
  Table~\ref{tab:phot_reg} and Fig.~\ref{fig:zoomvel} The left panel
  shows the entire observed wavelength range whereas the right one
  shows the spectral region around \mbox{\rm H$\alpha$} and
  {\mbox{[SII]$_{\lambda\lambda 6717,6731}$}}. \label{fig:spectra}}
  
\figcaption[RODRIGUES.fig04.ps]{Slit positions for the spectra of
  Table~\ref{tab:spec}. Slit lengths are not in
  scale. \label{fig:slit}}

\figcaption[RODRIGUES.fig06.ps]{\mbox{\rm H{\sc i}} contour map
  superimposed to the R image of NGC\,6845.  Full lines: \mbox{\rm
    H{\sc i}} associated to NGC\,6845\,A, with count peak flux =
  $4.08\times10^3 Jy/B*m/s$. Ten equally spaced contour levels from
  $0.4 \times10^3$ to $4.0 \times10^3 Jy/B*m/s$ are plotted. Dashed
  lines: \mbox{\rm H{\sc i}} associated to NGC\,6845\,B (mean velocity
  6700\,km\,s$^{-1}$, with count peak flux = $0.95 \times10^3
  Jy/B*m/s$. Contour levels are $0.4\times10^3$ and $0.8\times10^3
  Jy/B*m/s$. Beam size is shown at the lower left corner.
  \label{fig:r_hi}}

\figcaption[RODRIGUES.fig07.ps]{Rotation curve of NGC\,6845\,A.
  Different symbols are used to distinguish between SW (filled
  circles) and NE (empty triangles) directions. The full line gives
  our best Satoh model fit to the observed data set ($\chi^2=48.5$ and
  $RMS=18.1$; $a=6.5 \pm 0.4$ and $b/a=0.3 \pm 0.1$). Dashed lines
  give the upper and lower limits of the fitted curve within the
  observational errors.
  \label{fig:rota}}
  
\figcaption[RODRIGUES.fig08.ps]{\mbox{\rm H{\sc i}} velocity map
  superimposed to the R image of NGC\,6845.  Velocity levels are from
  6350 to $6700\rm{\,km\,s}^{-1}$ in intervals of
  $50\rm{\,km\,s}^{-1}$. \label{fig:r_vel}}

\figcaption[RODRIGUES.fig09.ps]{NGC\,6845\,A \mbox{\rm H{\sc i}}
  position--velocity diagram, obtained by cutting the data cube along
  the galaxy major axis, coincident with the slit number 4 in
  Figure~\ref{fig:slit}, the one used to obtain the optical rotation
  curve of NGC\,6845\,A (Fig.~\ref{fig:rota}). X--axis corresponds to
  the cut projection on the declination direction. Contour levels are
  drawn at 0.30, 0.45, 0.75, 1.0, and 1.5\,Kelvin. \label{fig:prof}}

\figcaption[RODRIGUES.fig10.ps]{Evolution of the encounter of model
  galaxies A (with gas) and C (without gas), viewed in the ``sky
  plane''. The top row shows only the stellar content, while the
  bottom row shows only gas particles. The frames measure 112\,kpc on
  a side, and time is shown in the upper right. \label{fig:simulac}}

\figcaption[RODRIGUES.fig11.ps]{Evolution of the encounter of galaxy
  model A and a point mass representing galaxy B, viewed in the ``sky
  plane''. The top row shows the stellar content, while the bottom row
  shows the gas particles. The frames measure 112\,kpc on a side, and
  time is shown in the upper right. \label{fig:simulab}}

\end{document}